\documentclass[10pt]{article}
\usepackage{amsmath}
\usepackage{amssymb}
\usepackage{fullpage}
\usepackage[dvips]{graphicx}
\usepackage[dvips]{epsfig}

\def\##1{{\bf #1}}
\def\=#1{\bar{#1}}

\def\eps{\epsilon}
\def\epso{\epsilon_0}
\def\muo{\mu_0}
\def\ko{k_0}
\def\lambdao{\lambda_0}
\def\etao{\eta_0}
\def\.{\mbox{ \tiny{$^\bullet$} }}

\def\le{\left(}
\def\ri{\right)}
\def\les{\left[}
\def\ris{\right]}
\def\lec{\left\{}
\def\ric{\right\}}

\def\r#1{(\ref{#1})}

\def\up{\hat{\bf{u}}_+}
\def\um{\hat{\bf{u}}_-}
\def\ux{\hat{\bf{u}}_x}
\def\uy{\hat{\bf{u}}_y}
\def\uz{\hat{\bf{u}}_z}

\def\aal{a_L}
\def\aar{a_R}
\def\bbl{r_L}
\def\bbr{r_R}
\def\ccl{t_L}
\def\ccr{t_R}

\def\Trr{ T_{RR} }

\def\Tll{T_{LL}  }

\def\Rrr{ R_{RR} }

\def\Rll{R_{LL}  }

\def\Ezdc{E_z^{dc}}

\begin{document}

\begin{center}
{\large {\bf Narrowband and ultranarrowband filters with
electro--optic structurally chiral materials}}

Akhlesh Lakhtakia

{\small \emph{CATMAS~---~Computational \& Theoretical Materials Sciences Group, \\
Department of Engineering Science \& Mechanics,\\
Pennsylvania State University, University Park, PA 16802--6812, USA.\\
Tel: +1 814 863 4319; Fax: +1 814 865 9974; E--mail: akhlesh@psu.edu}}
\end{center}
\noindent{\small
When a circularly polarized plane wave is normally incident
on a slab of a
structurally chiral material with local $\bar{4}2m$ point group symmetry and a
central twist defect, the slab can function as either a narrowband reflection hole filter
for co--handed plane waves
or an ultranarrowband transmission hole filter for cross--handed
plane waves, depending on its
thickness and the magnitude of the applied dc electric field. Exploitation of
the Pockels effect significantly reduces the thickness of the slab.
}\\

\noindent{\bf 1 Introduction}

Upon illumination by a normally incident, circularly polarized (CP) plane wave,
a slab of a
structurally chiral material (SCM) with its axis of nonhomogeneity aligned
 parallel to the thickness direction, is axially excited and reflects as well as
 transmits. Provided the SCM slab is
periodically nonhomogeneous and
sufficiently thick, and provided the wavelength of the incident plane wave lies in a
certain wavelength regime, the circular Bragg phenomenon is
exhibited. This phenomenon may be described as follows: reflection is
very high if the handedness of the plane wave is the same as the
structural handedness of the SCM, but is very low if the two handednesses
are opposite of each other. This phenomenon has been widely used
to make circular polarization filters of chiral liquid crystals [1]
and chiral sculptured thin films [2]. If attenuation with the
SCM slab is sufficiently low, it can thus function as
a CP
rejection filter. The circular Bragg phenomenon is robust enough that
periodic perturbations of the basic helicoidal nonhomogeneity can be altered to obtain 
different polarization--rejection characteristics [3]--[6].

In general, structural defects  in periodic materials produce
localized modes of wave resonance either within the Bragg regime or
at its edges. 
Narrowband CP
filters have been fabricated by incorporating either a
layer defect or a twist defect in the center of a SCM [2].
In the absence of the central defect, as stated earlier,
co--handed CP light is substantially reflected in the Bragg regime
while cross--handed CP light is not. The central defect creates a
narrow transmission peak for co--handed CP light that pierces the
Bragg regime, with the assumption that
dissipation in the SCM slab is negligibly small. 

Numerical simulations show that, as
the  total thickness of a SCM slab with a central defect increases, 
the bandwidth of the narrow
transmission peak begins to diminish and 
an even narrower
peak begins to develop in the reflection spectrum of the
cross--handed CP plane wave. 
There is a crossover thickness of the
device at which the two peaks are roughly equal in intensity.
Further increase in device thickness causes the co--handed
transmission peak to diminish more and eventually
vanish, while the cross--handed reflection
peak gains its full intensity and then saturates [7], [8]. The bandwidth of the 
cross--handed
reflection peak is a small fraction of that of the co--handed
transmission peak displaced by it. Such a crossover phenomenon 
cannot be exhibited by the
commonplace scalar Bragg gratings, and is unique to periodic
SCMs [8]. An explanation for the crossover phenomenon has recently
been provided in terms of coupled wave theory [9].

Although the co--handed transmission peak (equivalently, reflection hole)
has been observed and even utilized for both sensing [10] and lasing [11],
the cross--handed reflection peak (or transmission hole) remains entirely
a theoretical construct. The simple reason is that the total thickness for
crossover is very large [7]--[9]. Even a small amount of dissipation
evidently vitiates the conditions for
the emergence of cross--handed reflection peak. Clearly,
if the crossover thickness could be significantly reduced, the chances for
the development of the cross--handed reflection peak would be
greatly enhanced.

Such a reduction could be possible
if the SCM were to display the Pockels effect [12]~---~this thought 
emerged as a result of establishing the effect of a dc electric field on a
defect--free SCM endowed with a local $\bar{4}2m$ point group
symmetry [13]. A detailed investigation, as reported in the following 
sections, turned out to validate the initial idea. 

The plan of this paper is as follows: Section 2 contains a description
of the boundary value problem when a CP plane wave is normally incident
on a SCM slab with local $\bar{4}2m$ point group symmetry and a
central twist defect. Section 3
contains sample numerical results to demonstrate that  the chosen
device can function as either a narrowband reflection hole filter
or an ultranarrowband transmission hole filter~---~depending on (i) the
thickness of the SCM slab, (ii) the handedness of the incident plane wave,
and  (iii) the magnitude of the applied dc electric field.
Vectors are denoted in boldface; the  cartesian unit vectors are represented
by $\ux$, $\uy$, and $\uz$; symbols
for column vectors and matrixes
are decorated by an overbar;  and an $\exp(-i\omega t)$ time--dependence is
implicit with $\omega$ as the angular frequency.
The wavenumber and the intrinsic impedance of free space are denoted by $\ko=\omega\sqrt{\epso\muo}$ and
$\etao=\sqrt{\muo/\epso}$, respectively, with $\muo$ and $\epso$ being  the permeability and permittivity of
free space.\\

\noindent{\bf 2 Boundary Value Problem}

Suppose that a SCM slab with a central twist defect occupies the region
$0\leq z\leq 2L$,
the halfspaces $z \leq 0$ and
$z \geq 2L$ being vacuous.
An arbitrarily polarized plane wave is
normally incident on the device
from the halfspace $z \leq 0$.
In consequence, a reflected plane wave also exists in  the same halfspace
and a transmitted plane wave in the halfspace $z \geq 2L$.

The total electric field phasor  in the halfspace $z\leq 0$ is given by
\begin{equation}
\#{E}(\#{r}) =
 \le  \aal \, \up + \aar \, \um \ri \,
\exp( i \ko z ) + \le  \bbl \, \um + \bbr \, \up \ri \,
\exp( -i \ko z ) \, , \quad z \leq 0\, ,
\end{equation}
where $\#u_\pm = (\ux \pm i \uy)/\sqrt{2}$. Likewise,
the electric field phasor in the halfspace $z\geq 2L$ is
represented as
\begin{equation}
\#{E}(\#{r}) =
 \le  \ccl \, \up + \ccr \, \um \ri \,
\exp\les i \ko (z -2L)\ris  \,, \quad  z \geq 2L \, .
\end{equation} 
Here, $\aal$ and $\aar$ are the known amplitudes
of the left-- and the right--CP (LCP \& RCP)
components
of the incident plane wave;
$\bbl$ and $\bbr$ are the unknown amplitudes
of the reflected plane wave components; while $\ccl$ and $\ccr$
are the unknown amplitudes
of the transmitted plane wave components. The aim in solving the
boundary value problem is to determine $r_{L,R}$ and $t_{L,R}$
for known $\aal$ and $\aar$.

\noindent\emph{2.1 Electro--optic SCM with Local $\bar{4}2m$ Symmetry}

The chosen electro--optic SCM slab has the $z$ axis as its axis
 of chiral nonhomogeneity, and is subject to
a dc electric field ${\bf E}^{dc} = E_z^{dc}\,\uz$. The slab is assumed to have a
local $\bar{4}2m$ point group symmetry.

The optical relative permittivity matrix in the region
$0<z<2L$ may be stated  as follows [13]:
\begin{eqnarray}
\nonumber
&&\bar{\eps}^{SCM}(z) = \bar{S}_{z}\les
h\frac{\pi z}{\Omega}+h\psi(z)\ris\cdot\bar{R}_{y}(\chi)
\\[5pt]
\nonumber
&&\qquad\qquad
\cdot\left( 
\begin{array}{ccc}
\epsilon _{1}^{(0)} & \,-r_{63}\,\epsilon _{1}^{(0)2}\Ezdc\sin\chi & 
0 \\[5pt] 
-r_{63}\,\epsilon _{1}^{(0)2}\Ezdc\sin\chi & \,\epsilon _{1}^{(0)} & 
\,-r_{41}\,\epsilon _{1}^{(0)}\epsilon _{3}^{(0)}\Ezdc\cos\chi \\[5pt] 
0 & 
\,-r_{41}\,\epsilon _{1}^{(0)}\epsilon _{3}^{(0)}\Ezdc\cos\chi & \,\epsilon
_{3}^{(0)}
\end{array}
\right)
\\[10pt]
&&\qquad\qquad\qquad
\cdot\bar{R}_{y}(\chi)\cdot \bar{S}_{z}^{-1}\les
h\frac{\pi z}{\Omega}+h\psi(z)\ris\,,\quad 0< z< 2L\,.
\label{AAepsr}
\end{eqnarray}
Whereas $\epsilon _{1}^{(0)}$ and $\epsilon _{3}^{(0)}$ are, respectively,
the squares of the ordinary and the extraordinary refractive indexes
in the absence of the Pockels effect, $r_{41}$ and $r_{63}$ are the electro--optic
coefficients relevant to the $\bar{4}2m$ point group symmetry [12];
and only the lowest--order approximation of the Pockels effect has been
retained on the right side of \r{AAepsr}.
The tilt matrix 
\begin{equation}
\bar{R}_{y}(\chi )=\left( 
\begin{array}{ccc}
-\sin \chi & 0 & \cos \chi \\ 
0 & -1 & 0 \\ 
\cos \chi & 0 & \sin \chi
\end{array}
\right)
\end{equation}
involves the angle $\chi \in\les0,\pi/2\ris$ with respect to the $x$ axis in the
$xz$ plane. The use of the rotation matrix 
\begin{equation}
\bar{S}_z(\zeta)=\left( 
\begin{array}{ccc}
 \cos \zeta & -\,\sin\zeta&0 \\ 
\sin\zeta & \cos \zeta&0\\
0&0&1
\end{array}
\right)
\end{equation}
in \r{AAepsr}
involves the half--pitch $\Omega $ of the SCM along the $z$ axis. In
addition, the handedness parameter $h=1$ for structural right--handedness
and $h=-1$ for structural left--handedness. 

The angle $\psi(z)$ helps delineate the central twist as follows:
\begin{equation}
\psi(z)=
\left\{
\begin{array}{ll}
0\,,&\qquad 0<z<L\,\\
\Psi\,, &\qquad L <z<2L\,.
\end{array}\right.
\end{equation}
The angle $\Psi\in[0,\pi]$ is a measure of the central twist defect.

\noindent\emph{2.2 Reflectances and Transmittances}

The procedure to obtain the unknown reflection and transmission amplitudes
involves the 4$\times$4
matrix relation [2]
\begin{equation}
\label{eq8}
\=f_{exit} = \=M\cdot\=f_{entry}\,,
\end{equation}
 where the column 4--vectors
\begin{equation}
\label{eq9}
\=f_{entry} = \frac{1}{\sqrt{2}}\, \left(
\begin{array}{cc}
(\bbl+\bbr) + (\aal+\aar)\\ i\les -(\bbl-\bbr) + (\aal-\aar)\ris\\
-i\les (\bbl-\bbr) + (\aal-\aar)\ris/\etao\\
-\les (\bbl+\bbr) - (\aal+\aar)\ris/\etao \end{array}\right) \,
\end{equation}
and
\begin{equation}
\=f_{exit} = \frac{1}{\sqrt{2}}\, \left(
\begin{array}{cc}
 \ccl+\ccr\\ i\ (\ccl-\ccr) \\
-i (\ccl-\ccr)/\etao\\
(\ccl+\ccr) /\etao \end{array}\right)\,
\end{equation}
denote the electromagnetic fields at the entry and the exit
pupils, respectively.
The 4$\times$4 matrix
\begin{equation}
\label{meqn}
\=M =  \=B(h\Psi)\cdot\les\=B\left(
h\frac{\pi L}{\Omega}\right)\cdot\exp\left( i\=A^\prime L\right)\ris\cdot
\=B(-h\Psi)\cdot\les
 \=B\left(h\frac{\pi L}{\Omega}\right)
 \cdot\exp\left( i\=A^\prime L\right)\ris\,,
 \end{equation}
where
\begin{equation}
{\bar{A}}^\prime=\left(
\begin{array}{cccc}
0 & -\,\frac{ih\pi}{\Omega} & 0 &\omega\muo\\
\frac{ih\pi}{\Omega}& 0 &-\omega\muo & 0\\
-\omega\epso\eps_e &-\omega\epso\eps_1^{(0)}&0&-\,\frac{ih\pi}{\Omega}\\
\omega\epso\eps_d&\omega\epso\eps_e& \frac{ih\pi}{\Omega}&0
\end{array}
\right)\,,
\end{equation}
\begin{equation}
\bar{B}(\zeta)=\left( 
\begin{array}{cccc}
\cos \zeta & -\,\sin \zeta & 0 & 0 \\ 
\sin \zeta & \cos \zeta & 0 & 0 \\ 
0 & 0 & \cos \zeta &-\, \sin \zeta \\ 
0 & 0 & \sin \zeta & \cos \zeta
\end{array}
\right) \,,
\end{equation}
\begin{equation}
\epsilon _{d}=\frac{\epsilon _{1}^{(0)}\epsilon _{3}^{(0)}}{\epsilon
_{1}^{(0)}\cos ^{2}\chi +\epsilon _{3}^{(0)}\sin ^{2}\chi }\,,
\end{equation}
and
\begin{equation}
\epsilon _{e}=\Ezdc\,\epsilon _{1}^{(0)}\epsilon _{d}\left(
r_{41}\cos ^{2}\chi -r_{63}\sin ^{2}\chi \right)\,.
\end{equation}
The foregoing expression for $\=A^\prime$ is correct to the lowest order in both
$r_{41}\Ezdc$ and $r_{63}\Ezdc$.

The reflection amplitudes $r_{L,R}$ and the transmission
amplitudes $t_{L,R}$  can be computed for specified incident amplitudes
($\aal$ and $\aar$) by solving \r{eq8}.
 Interest usually lies  in determining
the reflection and transmission coefficients
entering the 2$\times$2 matrixes in the following two relations:
\begin{eqnarray}
\label{eq15}
\left( \begin{array}{cccc} \bbl \\ \bbr  \end{array}\right)  &=&
\left( \begin{array}{cccc} r_{LL} & r_{LR} \\ r_{RL} & r_{RR}\end{array}\right) \,
\left( \begin{array}{cccc} \aal \\ \aar  \end{array}\right) \, , \\
\label{eq16}
\left( \begin{array}{cccc} \ccl \\ \ccr  \end{array}\right)  &=&
\left( \begin{array}{cccc} t_{LL} & t_{LR} \\ t_{RL} & t_{RR}\end{array}\right) \,
\left( \begin{array}{cccc} \aal \\ \aar  \end{array}\right)
\, .
\end{eqnarray}
Both  2$\times$2 matrixes are defined phenomenologically.
The co--polarized transmission coefficients are denoted by $t_{LL}$ and
$t_{RR}$,
and the cross--polarized ones by $t_{LR}$ and $t_{RL}$; and similarly for the
reflection coefficients in \r{eq15}.
Reflectances and transmittances are denoted, e.g., as
$T_{LR} = |t_{LR}|^2$.\\

\noindent{\bf 3 Numerical Results}

Calculations of the reflectances and transmittances as functions
of the parameter $\lambdao/\Omega$ were made 
with and without electro--optic properties. The constitutive parameters used are that of
ammonium dihydrogen phosphate at $\lambdao=546$~nm [12], [14]:
$\eps_1^{(0)}=1.53^2$, $\eps_3^{(0)}=1.483^2$, $r_{41}=24.5\times 10^{-12}$~m~V$^{-1}$
and $r_{63}=8.5\times 10^{-12}$~m~V$^{-1}$. For illustrative results,
the SCM was chosen to be structurally right--handed (i.e., $h=1$) and the
tilt angle $\chi$ was fixed at $\pi/6$. The parameter $L/\Omega$ was constrained to
be an even integer.

\begin{center}
\begin{figure}[!htb]
\centering \psfull
\epsfig{file=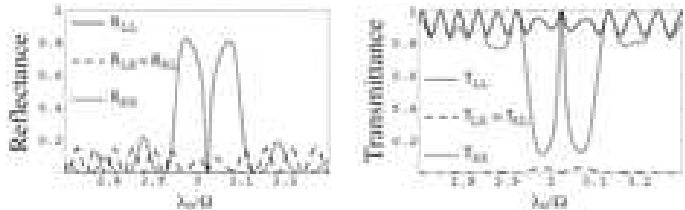, width=9cm}
%\bigskip
\caption{
{\small Reflectances ($R_{LL}$, etc.) and transmittances ($T_{LL}$, etc.) as functions
of the normalized wavelength $\lambdao/\Omega$, when $L= 30\Omega$, $
\Psi=90^\circ$, and $\Ezdc=0$. The other parameters are:
$\eps_1^{(0)}=1.53^2$, $\eps_3^{(0)}=1.483^2$, $r_{41}=24.5\times 10^{-12}$~m~V$^{-1}$, $r_{63}=8.5\times 10^{-12}$~m~V$^{-1}$, $h=1$, and $\chi=30^\circ$.
\label{Fig1}
}}
\end{figure}
\end{center}

\begin{center}
\begin{figure}[!htb]
\centering \psfull
\epsfig{file=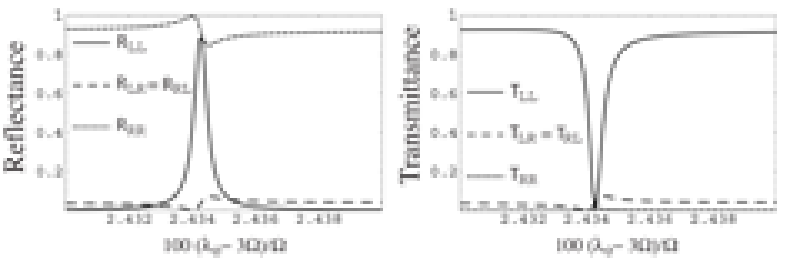, width=9cm}
%\bigskip
\caption{
{\small Same as Figure \ref{Fig1}, except that $L= 180\Omega$.
\label{Fig2}
}}
\end{figure}
\end{center}

Figures \ref{Fig1} and \ref{Fig2} present the variations of the reflectances and transmittances
with the normalized wavelength $\lambdao/\Omega$ when the Pockels
effect is not invoked (i.e., $\Ezdc=0$), for $L=30\Omega$ and $L=180\Omega$,
respectively.
The twist defect $\Psi=\pi/2$.
A co--handed reflection hole is clearly evident in the plot of $\Rrr$
at $\lambdao/\Omega\simeq 3.02$, and
the corresponding co--handed transmission peak may be seen in the
plot of $\Trr$ in Figure \ref{Fig1}. This hole/peak--feature is of high quality. As
the ratio $L/\Omega$ was increased, this feature began to diminish and
was replaced by a cross--handed transmission hole
in the plot of $\Tll$ along with a corresponding cross--handed peak
in the plot of $\Rll$. At $L=180\Omega$ (Figure \ref{Fig2}), the second feature is of
similar quality to the feature in Figure \ref{Fig1}. The bandwidth of the second feature is
a tiny fraction of the first feature, however. Neither of the two features
requires further discussion, as their distinctive features are known
well [7], [9], [15], [16], except to note that they constitute a defect mode of
propagation along the axis of chiral nonhomogeneity.

Figures 3 and 4 are the analogs of Figures 1 and 2, respectively,
when the Pockels effect has been invoked by setting $\Ezdc=1.5$~GV~m$^{-1}$.
Although $L=16\Omega$ in Figure 3, the narrowband feature therein is
of the same high quality as in Figure 1. The ultranarrowband
feature for
$L=58\Omega$
in Figure 4 is wider than its counterpart  in Figure 2, but could still be acceptable
for many purposes.
 The inevitable conclusion is that the incorporation of
the  Pockels effect in suitable SCMs provides a means
to realize thinner narrowband and ultranarrowband filters that are also
CP--discriminatory. This is the main result of this communication.

%%%%%%%%%%  Figure 1 begins %%%%%%%%%%%%
%%%%%%%%%%  Figure 1 ends  %%%%%%%%%%%%

%%%%%%%%%%  Figure 2 begins %%%%%%%%%%%%

%%%%%%%%%%  Figure 2 ends  %%%%%%%%%%%%

%%%%%%%%%%  Figure 3 begins %%%%%%%%%%%%
\begin{center}
\begin{figure}[!htb]
\centering \psfull
\epsfig{file=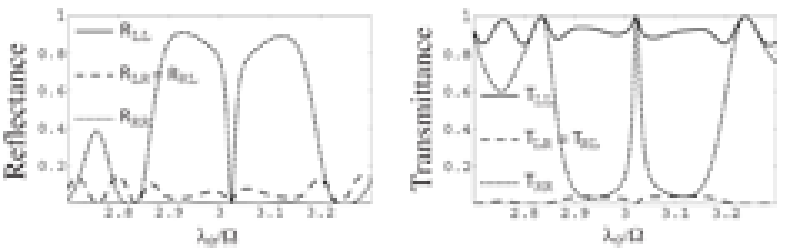, width=9cm}
%\bigskip
\caption{
{\small Same as Figure \ref{Fig1}, except that $L= 16\Omega$ and
$\Ezdc=1.5\times10^9$~V~m$^{-1}$.
\label{Fig3}
}}
\end{figure}
\end{center}

%%%%%%%%%%  Figure 3 ends  %%%%%%%%%%%%

%%%%%%%%%%  Figure 4 begins %%%%%%%%%%%%
\begin{center}
\begin{figure}[!htb]
\centering \psfull
\epsfig{file=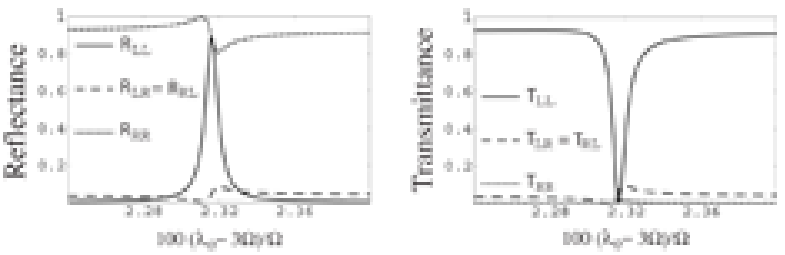, width=9cm}
%\bigskip
\caption{
{\small Same as Figure \ref{Fig2}, except that $L= 58\Omega$ and
$\Ezdc=1.5\times10^9$~V~m$^{-1}$.
\label{Fig4}
}}
\end{figure}
\end{center}

%%%%%%%%%%  Figure 4 ends  %%%%%%%%%%%%

An examination of the eigenvalues of $\=A^\prime$ shows that
the Bragg regime of the defect--free SCM is delineated by [13]
\begin{equation}
\lambda_{0_{min}}\leq {\lambdao} \leq
\lambda_{0_{max}}\,,
\label{Br-range}
\end{equation}
where
\begin{equation}
\lambda_{0_{min}}=2\Omega\,{\rm min}\lec \sqrt{\eps_{1\varphi}},\sqrt{\eps_{d\varphi}}\ric
\,,\end{equation}
\begin{equation}
\lambda_{0_{max}}=2\Omega\,{\rm max}\lec \sqrt{\eps_{1\varphi}},\sqrt{\eps_{d\varphi}}\ric
\,,
\end{equation}
\begin{equation}
\eps_{1\varphi}=\frac{1}{2}\,\les\eps_{1}^{(0)}+\eps_d +\,
\frac{\left(\eps_{1}^{(0)}-\eps_d\right)^2+4\eps_e^2}
{\eps_{1}^{(0)}-\eps_d}\,\cos2\varphi\ris\,,
\end{equation}
\begin{equation}
\eps_{d\varphi} =\frac{1}{2}\,\les\eps_{1}^{(0)}+\eps_d -\,
\frac{\left(\eps_{1}^{(0)}-\eps_d\right)^2+4\eps_e^2}
{\eps_{1}^{(0)}-\eps_d}\,\cos2\varphi\ris\,,
\end{equation}
and
\begin{equation}
\varphi=\frac{1}{2}\,\tan^{-1} \left(\frac{2h\eps_e}{\eps_d-\eps_1^{(0)}}\right)\,.
\end{equation}
Depending on the values of the constitutive parameters, the introduction
of $\Ezdc$ enhances the difference $\vert\sqrt{\eps_{d\varphi}}-\sqrt{\eps_{d\varphi}}
\vert$ significantly either for $\chi<\tan^{-1}\sqrt{ r_{41}/r_{63}}$ or
$\chi>\tan^{-1}\sqrt{ r_{41}/r_{63}}$. For the parameters selected for
Figures \ref{Fig1}--\ref{Fig4}, this enhancement is significant
for low values of $\chi$. The greater the enhancement, the faster does
the circular Bragg phenomenon develop as the normalized thickness
$L/\Omega$ is increased [2]. 
No wonder, the two types of spectral
holes appear for smaller values of $L/\Omega$ when $\Ezdc$ is
switched on.

%%%%%%%%%%  Figure 5 begins %%%%%%%%%%%%
\begin{center}
\begin{figure}[!h]
\centering \psfull
\epsfig{file=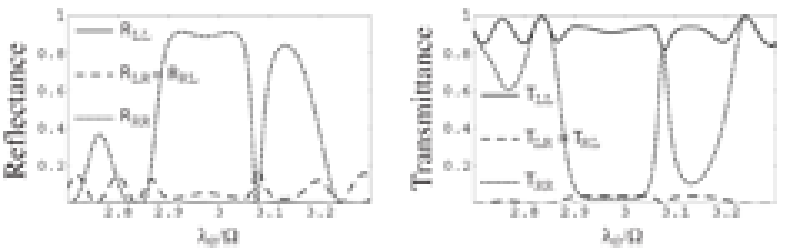, width=9cm}
\caption{
{\small Reflectances ($R_{LL}$, etc.) and transmittances ($T_{LL}$, etc.) as functions
of the normalized wavelength $\lambdao/\Omega$, when $L= 16\Omega$,
$\Psi=60^\circ$,
and $\Ezdc=1.5\times10^9$~V~m$^{-1}$. The other parameters are:
$\eps_1^{(0)}=1.53^2$, $\eps_3^{(0)}=1.483^2$, $r_{41}=24.5\times 10^{-12}$~m~V$^{-1}$, $r_{63}=8.5\times 10^{-12}$~m~V$^{-1}$, $h=1$, and $\chi=30^\circ$.
\label{Fig5}
}}
\end{figure}
\end{center}

%%%%%%%%%%  Figure 5 ends  %%%%%%%%%%%%

%%%%%%%%%%  Figure 6 begins %%%%%%%%%%%%
\begin{center}
\begin{figure}[!h]
\centering \psfull
\epsfig{file=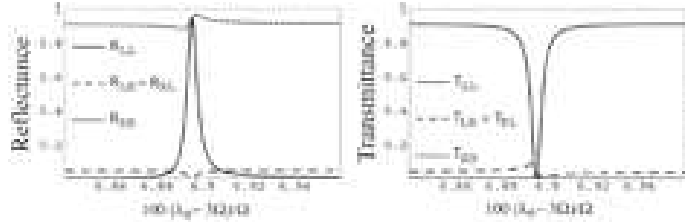, width=9cm}
\caption{
{\small Same as Figure \ref{Fig5}, except that
 $L= 70\Omega$.
\label{Fig6}
}}
\end{figure}
\end{center}

%%%%%%%%%%  Figure 6 ends  %%%%%%%%%%%%

Both types of spectral holes for $\Psi=\pi/2$ are positioned approximately
in the center of the wavelength regime \r{Br-range}, which is the Bragg regime
of a defect--free SCM [13]. For other values of
$\Psi$, the locations of the spectral holes may be estimated
as [9], [15]
\begin{equation}
\lambdao=\frac{1}{2}\left[\lambda_{0_{min}}+\lambda_{0_{max}} +
\left(\lambda_{0_{max}}-\lambda_{0_{min}}\right)\cos \Psi\right]\,.
\end{equation}
Figures \ref{Fig5} and \ref{Fig6} present sample results for
$\Psi=\pi/3$ in support.
However, let it be noted that the location of the spectral holes
can be manipulated simply by changing $\Omega$ while fixing
$\Psi=\pi/2$.\\

\noindent{\bf 4 Concluding Remarks}

The boundary value problem presented in this
paper is of the reflection and transmission of
a circularly polarized plane wave that is normally incident
on a slab of a
structurally chiral material with local $\bar{4}2m$ point group symmetry and a
central twist defect. Numerical results show that the slab can function as either a narrowband reflection hole filter for co--handed CP plane waves
or an ultranarrowband transmission hole filter for cross--handed
CP plane waves, depending on its
thickness and the magnitude of the applied dc electric field. Exploitation of
the Pockels effect significantly reduces the thickness of the slab for adequate
performance. The presented results are expected to urge experimentalists
to fabricate, characterize, and optimize the proposed devices.

\noindent\emph{\small This paper is affectionately dedicated to Prof. R. S. Sirohi
on the occasion of his retirement as the Director of the Indian Institute
of Technology, New Delhi.}\\

\smallskip
\noindent{\bf References}
{\small
\begin{enumerate}

\item
Jacobs S D (ed), \emph{Selected papers on liquid crystals for
optics.\/} (SPIE Optical Engineering Press, Bellingham, WA, USA), 1992.

\item
Lakhtakia A,  Messier R, \emph{Sculptured thin films:
Nanoengineered morphology and optics.\/} (SPIE Press, Bellingham, WA, USA), 2005,
Chap. 10.

\item
Polo Jr J A,  Lakhtakia A,
\emph{Opt. Commun.\/} 242 (2004) 13.

\item
Polo Jr J A,
\emph{Electromagnetics\/} 25 (2005) 409.

\item
Hodgkinson I, Wu Q h, De Silva L, Arnold M,  Lakhtakia A, McCall M,
\emph{Opt. Lett.\/} 30 (2005) 2629.

\item
Ross B M, Lakhtakia A, Hodgkinson I J, \emph{Opt. Commun.\/}
(doi:10.1016/j.optcom.2005.09.051).

\item
Kopp V I, Genack A Z, \emph{Phys. Rev. Lett.\/} 89 (2002) 033901.

\item
Wang F, Lakhtakia A, \emph{Opt. Commun.\/} 215 (2003) 79.

\item
Wang F, Lakhtakia A, \emph{Proc. R. Soc. Lond. A\/} 461 (2005) 2985.

\item
Lakhtakia A, McCall M W, Sherwin J A, Wu Q H,
Hodgkinson I J, \emph{Opt. Commun.\/} 194 (2002) 33.

\item
Schmidtke J, Stille W, Finkelmann H, \emph{Phys.
Rev. Lett.\/}   90 (2003) 083902.

\item
Boyd R W, \emph{Nonlinear optics.\/} (Academic Press, London, UK),
1992, Chap. 10.

\item
Reyes J A, Lakhtakia A,  \emph{Opt. Commun.\/}
(doi:10.1016/j.optcom.2005.08.034).

\item
http://www.kayelaby.npl.co.uk/general$_-$physics/2$_-$5/2$_-$5$_-$8.html

\item
Schmidtke J,   Stille W, \emph{Eur. Phys. J. E} 12 (2003)  553.

\item
Wang F, Lakhtakia A, \emph{Opt. Exp.} 13 (2005) 7319.

\end{enumerate}
}

\end{document}